\documentclass[prl,superscriptaddress,showpacs,preprint]{revtex4}
\usepackage{amssymb,amsmath,amsfonts}
\usepackage{epsfig}

\begin{document}

\title{Potential energy threshold for nano-hillock formation by impact of slow highly charged ions on a CaF$_2$(111) surface}
\date{\today}

\author{A.S. El-Said}
\affiliation{Institut f\"ur Allgemeine Physik, Vienna University of Technology, 1040 Vienna, Austria, EU}
\author{W. Meissl}
\affiliation{Institut f\"ur Allgemeine Physik, Vienna University of Technology, 1040 Vienna, Austria, EU}
\author{M.C. Simon}
\affiliation{Institut f\"ur Allgemeine Physik, Vienna University of Technology, 1040 Vienna, Austria, EU}
\author{J.R. Crespo L\'opez-Urrutia}
\affiliation{Max-Planck Institute for Nuclear Physics, 69029 Heidelberg, Germany, EU}
\author{C. Lemell}
\affiliation{Institute for Theoretical Physics, Vienna University of Technology, 1040 Vienna, Austria, EU}
\author{J. Burgd\"orfer}
\affiliation{Institute for Theoretical Physics, Vienna University of Technology, 1040 Vienna, Austria, EU}
\author{I.C. Gebeshuber}
\affiliation{Institut f\"ur Allgemeine Physik, Vienna University of Technology, 1040 Vienna, Austria, EU}
\author{HP. Winter}
\affiliation{Institut f\"ur Allgemeine Physik, Vienna University of Technology, 1040 Vienna, Austria, EU}
\author{J. Ullrich}
\affiliation{Max-Planck Institute for Nuclear Physics, 69029 Heidelberg, Germany, EU}
\author{C. Trautmann}
\affiliation{Gesellschaft f\"ur Schwerionenforschung (GSI), 64291 Darmstadt, Germany, EU}
\author{M. Toulemonde}
\affiliation{Centre Interdisciplinaire de Recherches Ions Laser (CIRIL), Laboratoire commun CEA,CNRS,UCBN,ENSICAEN,BP5133, 14070 Caen Cedex 5, France, EU}
\author{F. Aumayr}
\thanks{to whome correspondence should be addressed.}
\email[e-mail: ]{aumayr@iap.tuwien.ac.at}
\affiliation{Institut f\"ur Allgemeine Physik, Vienna University of Technology, 1040 Vienna, Austria}

\begin{abstract}
We investigate the formation of nano-sized hillocks on the (111) surface of CaF$_2$
single crystals by impact of slow highly charged ions. Atomic force microscopy reveals a
surprisingly sharp and well-defined threshold of potential energy carried into the
collision of about 14 keV for hillock formation. Estimates of the energy density
deposited suggest that the threshold is linked to a solid-liquid phase transition
(``melting'') on the nanoscale. With increasing potential energy, both the basal
diameter and the height of the hillocks increase. The present results reveal a
remarkable similarity between the present predominantly potential-energy driven process and
track formation by the thermal spike of swift ($\sim$ GeV) heavy ions.
\end{abstract}

\maketitle

The surface topography of many solids experiences drastic modifications when exposed to energetic ions. The changes induced depend on the target material as well as on various beam parameters such as charge, energy, mass and fluence of the incoming projectiles and can result in well-ordered patterns, such as ripples or self ordered dots \cite{ripp1,ripp2,ripp3}. Impact of single ions has been demonstrated to induce nano-sized hillocks on metals, semiconductors and dielectric targets. Remarkably enough, the hillocks observed have a similar height (a few nm) and diameter (20--40 nm) although the beam energies used span several orders of magnitude reaching up to GeV.

Impact of swift heavy ions is known to induce physical, chemical, and structural modifications not only on the surface but also in the bulk (see e.g.\ \cite{ref1,ref2,ref3} and Refs.\ therein). Individual projectiles form cylindrical tracks around their trajectory of a few nanometers in diameter. Track formation sets in above a critical value of the energy loss $dE/dx$ of the projectiles and occurs particularly in insulators (e.g.\ polymers, oxides, ionic crystals). Depending on the solid, tracks consist of amorphised or defect-rich material. In non-amorphisable alkali and alkaline earth halides (e.g.\ LiF and CaF$_2$) the damage process is governed by exciton-mediated defects such as color centers and defect clusters \cite{Trautmann98,Schwartz04}. Above a critical value of $dE/dx$, damage produced in the core of the track leads to a macroscopic volume increase (swelling, \cite{Boccanfuso02,Trautmann00}), track etchability \cite{Trautmann98}, and stress \cite{Manika03}. At the surface of ionic crystals, swift ions induce nanometric hillocks \cite{Khalfaoui05,El-Said04} above a threshold value similar to that for swelling \cite{Trautmann00}. Although numerous experimental data are available for hillock formation due to swift heavy ion impact, the principle of the mechanism is still not yet fully understood.

In this letter we present experiments with \textit{slow} ($v_p\approx 0.3$ a.u.) highly charged ions (HCI) which also induce hillock-like nanostructures on the surface of CaF$_2$ single crystals. These nanostructures closely resemble those created by fast ions. Moreover, we find a strong dependence of the formation on the potential energy rather than on the stopping power. Most surprisingly, we find a well-defined threshold of potential energy required for the onset of nano-hillock formation. Since CaF$_2$ is used as an insulator in silicon microelectronic devices \cite{smith,schowalter} epitaxially grown on semiconductor surfaces \cite{lucas}, our findings might be of importance for high resolution patterning of thin CaF$_2$ films on Si and for the creation of nanostructured templates for adlayer growth during fabrication of CaF$_2$/Si-based epitaxial insulator-semiconductor structures.

Our experiments were performed on air-cleaved CaF$_2$(111) surfaces. Cleavage is known to result in a fluorine-terminated surface. Contact-mode atomic force microscopy (AFM) in air revealed large atomically flat surfaces with occasional cleavage steps separating individual terraces. Several freshly cleaved CaF$_2$ samples were mounted in a vacuum chamber of pressure in the 10$^{-10}$ mbar range and irradiated normal to the (111) surface with HCI of kinetic energy below 5 keV per nucleon. The irradiation was performed at the Heidelberg electron beam ion trap \cite{ref6} using $^{40}$Ar$^{q+}$ ($q = 11$, 12, 14, 16, 17, and 18) as well as $^{129}$Xe$^{q+}$ ($q = 22$, 26, 28, 30, 33, 36, 40, 44, 46, and 48) projectiles during several runs. The extraction voltage was 10 kV (for Xe$^{44+}$ also 6.4 kV) equivalent to a kinetic energy of 10 kV (6.4 kV) times charge $q$ resulting in a projected range between 90 and 140 nm in CaF$_2$, assuming that stopping power and range are unaffected by the high charge state (see below) \cite{zbl}. The beam flux varied between $10^3$ and several 10$^4$ ions/s, and was measured via electron emission statistics with close to 100\% detection efficiency \cite{Meissl,Arnau}. After exposure to fluences up to $(0.5 - 5) \times 10^9$ ions/cm$^2$, the surface of the crystals was inspected in ambient air by contact-mode AFM. As reported earlier for CaF$_2$ single crystals irradiated with swift heavy ions, the surface hillocks are stable in atmosphere at room temperature \cite{Khalfaoui05}.

\begin{figure}
\centerline{\epsfig{file=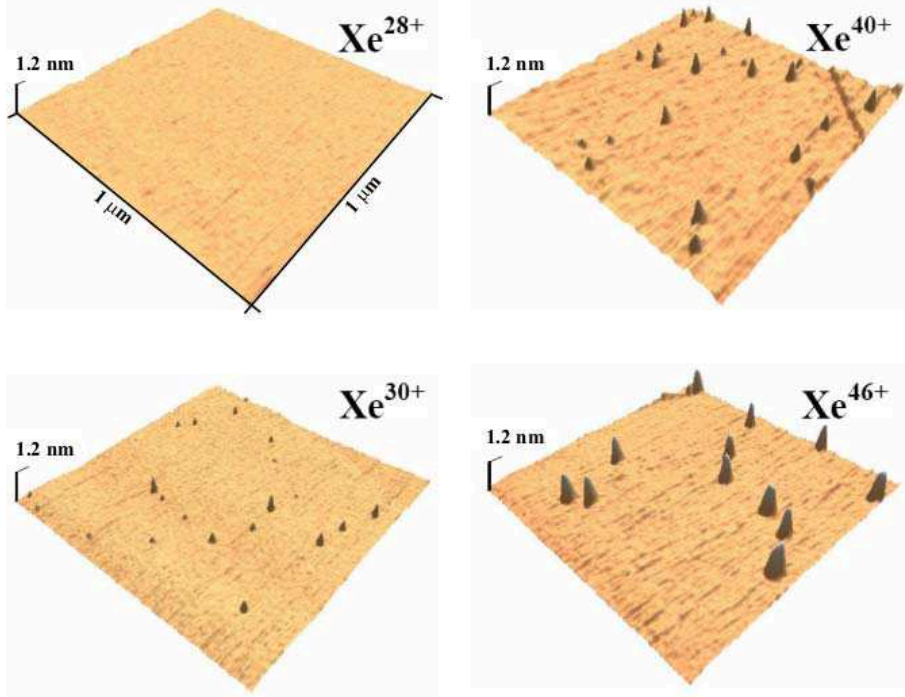,width=12cm}}
\caption{(Color online) Topographic contact-mode AFM images of a CaF$_2$(111) surface irradiated by $10q$ keV Xe$^{q+}$ ions of charge state $q=28$, 30, 40, 46. In each frame an area of $1 \mu$m $\times 1\mu$m is displayed. Hillock-like nanostructures protruding from the surface are only observed for Xe projectiles with charge state $q \geq 30$. Above this threshold, the height and diameter of the hillocks increase with ion charge state.}
\label{fig1}
\end{figure}

Fig.\ \ref{fig1} shows examples of AFM topographic images of CaF$_2$(111) after irradiation with Xe$^{28+}$ (2.2 keV/amu), Xe$^{30+}$ (2.3 keV/amu), Xe$^{40+}$(3.1 keV/amu), and Xe$^{46+}$ (3.6 keV/amu) ions. Hillock-like nanostructures protruding from the surface are observed for highly charged Xe$^{q+}$ ($q \geq 30$) and fully stripped Ar$^{18+}$ ions whereas targets irradiated with Xe$^{q+}$ ($q \leq 28$) and Ar$^{q+}$ ($q \leq 17$) projectiles did not exhibit any hillocks. The sharp transition, e.g.\ between $q=17$ and 18 of argon cannot be associated with irradiation parameters in an obvious way. Moreover, results from measurements with $6.4\cdot q$ keV (2.2 keV/amu) Xe$^{44+}$ differ by less than 5\% from the data of $10\cdot q$ keV (i.e.\ 3.4 keV/amu) Xe$^{44+}$ ions. It appears that the kinetic energy plays no decisive role for the size of the observed nanostructures.

\begin{figure}
\centerline{\epsfig{file=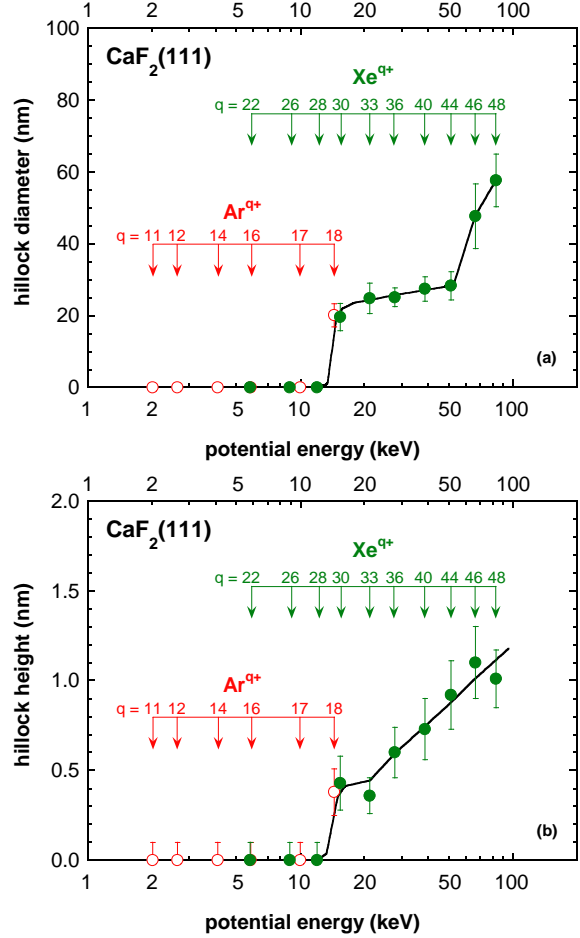,width=8cm}}
\caption{(Color online) Mean diameter (top) and height (bottom) of hillock-like nanostructures as a function of the potential energy of Ar$^{q+}$ (open symbol) and Xe$^{q+}$ (full symbol) projectiles. Hillocks are found only above a potential energy threshold of about 14 keV. The error bars correspond to the standard deviation of the diameter and height distributions; the solid lines are drawn to guide the eye.}
\label{fig2}
\end{figure}

The AFM images were evaluated with respect to number density, height and width distributions of the hillocks. The number of the hillocks per unit area was found to be in good agreement with the applied ion fluence, i.e., above the threshold, a large majority of projectiles ($>70$\%) produces an individual hillock. Their height ranges between 0.5 and 1 nm and their diameter between $\sim 20$ and 60 nm. Due to the finite curvature radius of the AFM tip (nominally 4-5 nm), the diameter (but not the height) of the hillocks is subject to a systematic error. The protrusions are rather flat with a diameter to height ratio between 40 and 60. In contrast to hillocks induced by swift heavy ions \cite{Khalfaoui05}, we observe only a weak correlation between the diameter and height value of a given hillock. Furthermore, the size data were found to be strongly dependent on the potential energy the projectile carries into the HCI-surface collision (Fig.\ \ref{fig2}).

The potential (i.e.\ internal) energy $E_p$ of HCI is equal to the total ionization energy required for producing the high charge state from its neutral ground state. $E_p$ is known to have a strong influence on surface interaction processes such as electron emission, sputtering, and secondary ion emission \cite{Arnau}. For both Xe and Ar ions a remarkably well-defined sharp threshold in potential energy (between $E_p\approx 12$ keV for Xe$^{28+}$ and $E_p\approx 14.4$ keV for Ar$^{18+}$) for hillock formation emerges. Above this threshold, an increase of the potential energy leads to an increase of both the basal diameter and the height of the hillocks. Another steep increase of the mean hillock diameter potentially indicating a second threshold is found between Xe$^{44+}$ and Xe$^{46+}$ (top of Fig.\ \ref{fig2}). 

A convenient starting point for an analysis of the observed hillock formation is the interaction of the HCI above the surface involving a series of complex processes on different time and energy scales. When the ion approaches the surface, neutralization starts by electron transfer from the target into highly excited states of the projectile \cite{burg,hagg,wirtz}. Deexcitation of the projectile proceeds via Auger-type processes producing primarily low energy electrons. Only for very highly charged heavy ions with open K and L shells electron energies up to several keV can be expected. For these states, however, radiative decay becomes important as a competing deexcitation mechanism with fluorescence yields of typically $\sim 12$\% \cite{bhalla}. An increasing amount of potential energy is therefore dissipated by X-ray emission. The critical distance $R_c$ from the surface for electron transfer to the HCI can be estimated as \cite{barany}
\begin{equation}
R_c\approx\frac{\sqrt{2q\varepsilon(8i+\varepsilon-1)}}{(\varepsilon+1)W}\, ,
\end{equation}
where $i$ is the amount of charge left behind (for the first electron capture $i=1$), and $W$ and $\varepsilon$ are the workfunction and the dielectric constant of the material, respectively. For CaF$_2$ we find $R_c\approx 0.16\sqrt{q}$ nm, which sets an upper limit for the time available for the above-surface neutralization sequence. As an example, for an ion of $q=40$, $R_c$ is about 1 nm and the neutralization time is of the order of 1 fs. As the projectile velocity is also proportional to $\sqrt{q}$ in our experiment, the above-surface interaction time is the same for all projectiles with equal acceleration voltage. The transfer of electrons to the projectile leaves unbalanced holes in the surface which store part of the potential energy of the HCI. It is known from electron-emission yield measurements that $\sim 3\, q$ electrons are emitted per projectile \cite{Arnau}. For the impact of a $q=40$ ion we therefore estimate a number of about 150 unbalanced holes (emitted electrons + electrons required for neutralization) created. They diffuse only slowly into the material (hole velocity in the valence band derived from tight-binding calculations is smaller than 0.33 nm/fs \cite{albert}). Furthermore, two holes (F$^0$ atoms) in adjacent sites recombine to volatile fluorine gas molecules leaving behind a Ca-enriched metallic surface. Upon impact of the projectile the target is structurally weakened and features fluorine depleted, defect-enriched areas.

For an analysis and interpretation of our data, we adapt aspects of the inelastic thermal spike model developed for swift ions \cite{ref8}. The underlying assumption is that the initial deposition of projectile energy involves the electronic subsystem of the target and proceeds on a (sub-) femtosecond scale while the energy transfer to the lattice and the concomitant lattice deformation and defect production occurs on a (sub-) picosecond scale. The present case of slow HCI differs, however, substantially in two aspects. The primary energy distribution of ``hot'' electrons results from a relaxation process of a hollow atom rather than from a Bethe-Born type ionization spectrum. A significant fraction of the potential energy is emitted by energetic ($\sim$ keV) inner-shell Auger electrons. Moreover, slow HCI deposit their potential energy in a shallow surface region, whereas swift ions deposit kinetic energy along the full length of their trajectory within a cylindrical volume.

In the following we estimate the amount of energy and the target volume in which HCI deposit their potential energy. From calorimetric measurements it is known \cite{Kentsch} that only part of the potential energy is transferred to the target. We suppose that this fraction is around 50\% with an uncertainty of  20\%. The excited target electrons spread their energy within $\sim 100$ fs by diffusion into a hemispherical volume around the impact site with a radius large compared to the source volume determined by the hollow-atom relaxation ($R_c\approx 1$ nm). In turn, the diffusing hot electron distribution transfers energy to the lattice by inelastic electron-phonon scattering with a characteristic time constant $\tau_{e-ph}$ of a few hundred fs. Phonon-mediated energy transport leads to further spread and thermalization.

Assuming, for simplicity, the same overall thermal diffusion length $\lambda_D\approx 4$ nm as observed for swift ions in CaF$_2$ \cite{ref8}, the fraction of internal energy $E_D$ is deposited in a hemisphere of radius $\lambda_D$ comprising about $N\approx 8.5\times 10^3$ atoms. If the energy deposition per atom, $E_D/N$, within this locally heated volume exceeds the melting energy of $E_M=0.55$ eV/atom \cite{NN6} a solid-liquid phase transition is expected. Likewise, for $E_D/N>E_S=1.55$ eV/atom \cite{NN6}, sublimation should set in. In order to have these energies available at the impact site, the HCI needs a potential energy above $E^{th}_M=14$ keV and $E^{th}_S=40$ keV, respectively. Such a crude estimate carries a large error bar of about $\pm 50$\% due to the uncertainty in the effective $\lambda_D$ and the fraction of deposited energy. The estimates are remarkably, maybe even fortuitously, close to the observed threshold for hillock formation ($E^{th}_M$) and for the second drastic, almost steplike size increase ($E^{th}_S$).

It should be noted, that the conceptual difficulty in applying the model of the thermalization of the internal energy within $\lambda_D$ to slow HCI lies in the fact that the difference in internal energy between subthreshold (Ar$^{17+}$) and above threshold (Ar$^{18+}$) is emitted in one additional K-Auger electron with an energy of $E_K\approx 4.5$ keV and its large inelastic mean free path $\lambda_K\gg\lambda_D$. The deposition of this energy difference is thus not confined to the critical volume of melting or evaporation. Moreover, a large fraction of K-Auger electrons emitted near the surface is directly ejected into vacuum and a fraction of K-holes is dissipated by X-rays and thus unavailable for thermalization.

An alternative and additional heating mechanism could be the pre-equilibrium charge-state dependent electronic and nuclear stopping in insulators strongly deviating from standard values for near-neutral projectiles in equilibrium \cite{schenkel}. Such deviations have been found for low but not negligible projectile velocity of $v_p\approx 0.3\, v_{Bohr}$. In a shallow region at and below the surface, a strong enhancement with charge state $q$ of the kinetic energy deposition and, correspondingly, reduction of range is expected for highly charged ions. This could increase the energy deposition $dE/dx$ near the surface to above the critical value for phase transition observed for swift heavy ions \cite{ref8}. Future experiments at lower $v_p$ should shed light on the role of this energy deposition process.

Irrespective of the not yet fully understood details of the heating mechanism, the following scenario emerges: hillock formation is the result of local melting and swelling when the energy deposition by HCI near the surface exceeds the melting energy $E_M=0.55$ eV/atom. If the energy deposition exceeds the critical value for sublimation $E_S=1.55$ eV/atom, evaporation should lead to the formation of blisters of enhanced size. Moreover, one should expect the transition from blister to crater formation when the evaporation is further enhanced. This scenario, however, suggests that crater formation should be more likely for even higher $q$ and at near-grazing impact angles when the energy deposition concentrates near the topmost atomic layer, and direct evaporation into vacuum becomes possible.

In conclusion, the bombardment of a CaF$_2$ surface with moderately slow ($v_p=0.3$ a.u.) highly charged Ar and Xe ions produces hillock-like surface nanostructures. The formation of these protrusions requires a critical potential energy of 14 keV (Ar$^{18+}$ and Xe$^{30+}$). A second threshold characterized by a steep increase of hillock diameter appears at 50 keV (Xe$^{44+}$). In analogy to hillock formation by swift heavy ions, we associate the two thresholds with phase transitions of melting and sublimation caused by the deposition of the potential energy within the electronic subsystem. The presently discussed scenario suggests future investigations of HCI induced nanostructures at smaller $v_p$, larger $q$, and grazing incidence.

This work has been supported by Austrian Science Foundation FWF (Projects No.\ 17449 and
M894-N02). The irradiation experiments
were performed at the distributed LEIF-Infrastructure at MPI Heidelberg Germany, supported
by Transnational Access granted by the European Project HPRI-CT-2005-026015.


\begin{thebibliography}{25}
\bibitem{ripp1}S. Facsko et al., Science {\bf 285} 1551 (1999). 
\bibitem{ripp2}M. Castro, R. Cuerno, L. V\'azquez, and R. Gago, \prl {\bf 94}, 016102 (2005).
\bibitem{ripp3}B. Ziberi, F. Frost, Th. H\"oche, and B. Rauschenbach, \prb {\bf 72}, 235310 (2005).
\bibitem{ref1} R.L. Fleischer, P.B. Price, and R.M. Walker, J. Appl.\ Phys.\
{\bf 36}, 3645 (1965).
\bibitem{ref2} H. Dammak et al., \prl {\bf 74}, 1135 (1995).
\bibitem{ref3} L.T. Chadderton, Radiation Measurements {\bf 36}, 13 (2003).
\bibitem{Trautmann98} C. Trautmann, K. Schwartz, and O. Geiss, Journal of Appl.\ Phys.\ A {\bf 66}, 3560 (1998).
\bibitem{Schwartz04} K. Schwartz et al., Phys.\ Rev.\ B {\bf 70}, 184104 (2004); C. Trautmann, M. Toulemonde, K. Schwartz, J.M. Costantini, and A. M\"uller, Nucl.\ Instr.\ Meth.\ B {\bf 164-165} 365 (2000).
\bibitem{Boccanfuso02} M. Boccanfuso, A. Benyagoub, K. Schwartz, C. Trautmann, and M. Toulemonde, Nucl.\ Instr.\ Meth.\ B {\bf 191}, 301 (2002).
\bibitem{Trautmann00} C. Trautmann, M. Toulemonde, J.M. Costantini, J.J. Grob, and K. Schwartz, \prb {\bf 62}, 13 (2000); M. Boccanfuso, A. Benyagoub, K. Schwartz, M. Toulemonde, and C. Trautmann, Progr.\ Nucl.\ Energy {\bf 38}, 271 (2001).
\bibitem{Manika03} I. Manika, J. Maniks, K. Schwartz, M. Toulemonde, and C. Trautmann, Nucl.\ Instr.\ Meth.\ B {\bf 209}, 93 (2003).
\bibitem{Khalfaoui05} N. Khalfaoui et al., Nucl.\ Instr.\ Meth.\ B {\bf 240}, 819 (2005).
\bibitem{El-Said04} A.S. El-Said et al., Nucl.\ Instr.\ Meth.\ B {\bf 218}, 492 (2004); C. M\"uller et al., Nucl.\ Instr.\ Meth.\ B {\bf 209}, 175 (2003). 
\bibitem{smith} T.P. Smith, J.M. Phillips, W.M. Augustyniak, and P.J. Stils, Appl.\ Phys.\ Lett.\ {\bf 45}, 907 (1984).
\bibitem{schowalter} L.J. Schowalter and R.W. Fathauer, J. Vac.\ Sci.\ Technol.\ {\bf 4}, 1026 (1986).
\bibitem{lucas} C.A. Lucas and D. Loretto, Appl.\ Phys.\ Lett.\ {\bf 60}, 2071 (1992).
\bibitem{ref6} J.R. Crespo L\'opez-Urrutia et al., Hyperfine Interactions {\bf 146/147},
109 (2003).
\bibitem{zbl} J.F. Ziegler, J.P, Biersack, and U. Littmark, ``The Stopping and Range of Ions in Solids,'' (Pergamon Press, New York, 1984).
\bibitem{Meissl}W. Meissl et al., Rev.\ Sci.\ Instrum.\ {\bf 77},(2006) in print.
\bibitem{Arnau}A. Arnau et al., Surf.\ Sci.\ Rep.\ {\bf 27}, 113 (1997).
\bibitem{burg} J. Burgd\"orfer, P. Lerner, and F.W. Meyer, \pra {\bf 44}, 5674 (1991).
\bibitem{hagg} L. H\"agg, C.O. Reinhold, and J. Burgd\"orfer, \pra {\bf 55}, 2097 (1997).
\bibitem{wirtz} L. Wirtz, C.O. Reinhold, C. Lemell, and J. Burgd\"orfer, \pra {\bf 67}, 12903 (2003).
\bibitem{bhalla} C.P. Bhalla, \pra {\bf 8}, 2877 (1973).
\bibitem{barany} A. Barany and C.J. Setterlind, Heavy Ion Phys.\ {\bf 1}, 115 (1995).
\bibitem{albert} J.P. Albert, C. Jouanin, and C. Gout, \prb {\bf 16}, 925 (1977).
\bibitem{ref8} M. Toulemonde, Ch. Dufour, A. Meftah, and E. Paumier, Nucl.\ Instr.\ Meth.\ B {\bf 166/167}, 903 (2000); A. Meftah et al., Nucl.\ Instr.\ Meth.\ B {\bf 237}, 563 (2005).
\bibitem{Kentsch} U. Kentsch, H. Tyrroff, G. Zschornack, and W. M\"oller, \prl {\bf 87}, 105504 (2001).
\bibitem{NN6} G. Grochtmann, R.J. Meyer, F. Peters, and L. Gmelin, ``Gmelins Handbuch der anorganischen Chemie,'' (Verlag Chemie, Berlin, 1970).
\bibitem{schenkel} T. Schenkel et al., \prl {\bf 79}, 2030 (1997).
\end{thebibliography}
\end{document}